\documentstyle{l-aa}

\font
\capt=cmti8.tfm    
\def\lsim{\lower.5ex\hbox{$\; \buildrel < \over \sim \;$}}
\def\gsim{\lower.5ex\hbox{$\; \buildrel > \over \sim \;$}}
\begin{document}
\title { X-ray spectral components in the hard state of GRS 1915+105 :
origin of the 0.5 -- 10 Hz QPO }
\author{A.R. Rao$^1$, S. Naik$^1$, S.V. Vadawale$^1$ and  Sandip K. Chakrabarti$^2$ }
\institute { $^1$Tata Institute of Fundamental Research, Homi Bhabha Road, Mumbai(Bombay) 400 005, India\\
$^2$S.N. Bose National Centre for Basic Sciences, Salt Lake, Calcutta 7000091, India}
\offprints { A.R. Rao  {\it arrao@tifr.res.in}}
\date{Received ; accepted  }
\thesaurus{13.25.5; 02.01.2; 02.02.1; 08.23.3; 08.9.2 GRS 1915+105}
\maketitle
\markboth{Rao et al: X-ray spectral components in GRS 1915+105}{}

\maketitle

\begin{abstract}
We investigate the origin of the ubiquitous 0.5 -- 10 Hz QPO in the
Galactic microquasar GRS~1915+105. Using the archival X-ray data
from RXTE, we make a wide band X-ray spectral fitting to the
source during a low-hard state observed in 1999 June. 
We resolve the X-ray
spectra into three components, namely a multi-color disk 
component, a Comptonised component and a power-law at higher
energies. This spectral description is favored compared to other
normally used spectra like a cut-off power law, hard components with
reflection etc. We find that the 0.5 -- 10 Hz QPO is predominantly due to
variations in the Comptonised component. We use this result to
constrain the location of the various spectral components in the
source. 
\end{abstract}

\medskip

\section{Introduction}

The Galactic microquasar GRS~1915+105 is a bright X-ray source
and it is a subject of intense study in all wavelengths
(see Mirabel \& Rodriguez 1999, and references therein).
It has been exhibiting a multitude of types of X-ray variability
characteristics ranging from a quasi-stable high frequency QPO 
at 67 Hz (Morgan, Remillard,  \& Greiner 1997) to
long durations of stable emission modes (Belloni et al. 2000).
One of the interesting features of this source is the 
detection of a stable, narrow, and intense QPO in the
frequency range of 0.5 -- 10 Hz (Agrawal et al. 1996; Morgan
\& Remillard 1996). Chen, Swank, \& Taam (1997) found that this  QPO
emission is a   characteristic feature of the hard branch and it 
is absent in the soft branch.  Paul et al.  (1998) found 
that the 0.5 -- 10 Hz QPO traces
the change of state from a ``flaring state'' to a low-hard state quite
smoothly along with other X-ray characteristics like the low frequency
variability. Trudolyubov, Churazov, \& Gilfanov (1999b) studied the 
low state and state transitions using the RXTE data and
concluded that the QPO centroid frequency is correlated with the spectral
and timing parameters.
Muno, Morgan, \& Remillard (1999) 
found that the 0.5 -- 10 Hz QPO can be used as a tracer of the spectral
state of the source and the source mainly stays in two states: the spectrally
hard state with the QPO and the soft state without the QPO.
Another interesting feature of the 0.5  -- 10 Hz QPO is that it is also
found during the brief episodes of `off' states when the source
teeters between `off' and `on' states (Belloni et al. 1997;
Yadav et al. 1999).

 The large rms amplitude (about 10\%), narrow width and the relative stability 
over considerably long durations makes it difficult to explain 
the 0.5 -- 10 Hz QPO on the basis of the inner disk oscillations in the 
accretion disk. 
Chakrabarti \& Manickam (2000) showed that even in the spectrally
harder states, not all the photons
participated in QPOs. They found that, 0-4 keV soft photons showed little
or no sign of QPO while 4-13 keV
photons exhibited QPO. They interpret this as due to the oscillation of
the
region responsible for the hard radiation, i.e., the post-shock region.
They also detected a correlation  between the average frequency of the
QPO and the duration of the `off' states. Similar correlation with
centroid frequencies were
also found by Trudolyubov, Churazov, \& Gilfanov (1999a).

In this {\it Letter}, we perform a detailed X-ray spectroscopy of the source
using the RXTE data
archives. We find a better way to quantify the nature of the photons
which participated in QPOs.
We show that the entire spectrum can be split into a disk blackbody
component,
a thermal Compton component and an additional hard component and
only the Comptonised
component participates in QPOs.

\section{Analysis and results }

\subsection{Data selection}

The 0.5 $-$ 10  Hz QPO is detected during the low hard states as well as
during the low state of bursts (Muno et al. 1999). The long duration hard 
states are classified as the $\chi$ state in 
 Belloni et al. (2000), who have made a comprehensive study of RXTE PCA 
observations on GRS~1915+105 using the total count rates and color-color
diagrams. The $\chi$ state is further divided into $\chi_1$ to $\chi_4$ states
depending on the count rate and color. The $\chi_3$ state is also 
found to be associated with high radio emission - the ``plateau state'' 
(Fender et al. 1999).  
Each of these steady hard states stays for a long duration (a few days to 
a few months - see Belloni et al. 2000) and during a given steady state
the spectral and temporal behaviors are similar. To quantify the
relation between the QPO emission and the spectral parameters, we have
selected one observation in the $\chi_3$ state of the source, obtained
on 1999 June 7. A detailed
spectral and temporal analysis of all the steady hard states
of the source will be presented elsewhere (Vadawale et al. 2000).

\begin{figure}[t]
\capt
\vskip 7.0cm
\includegraphics{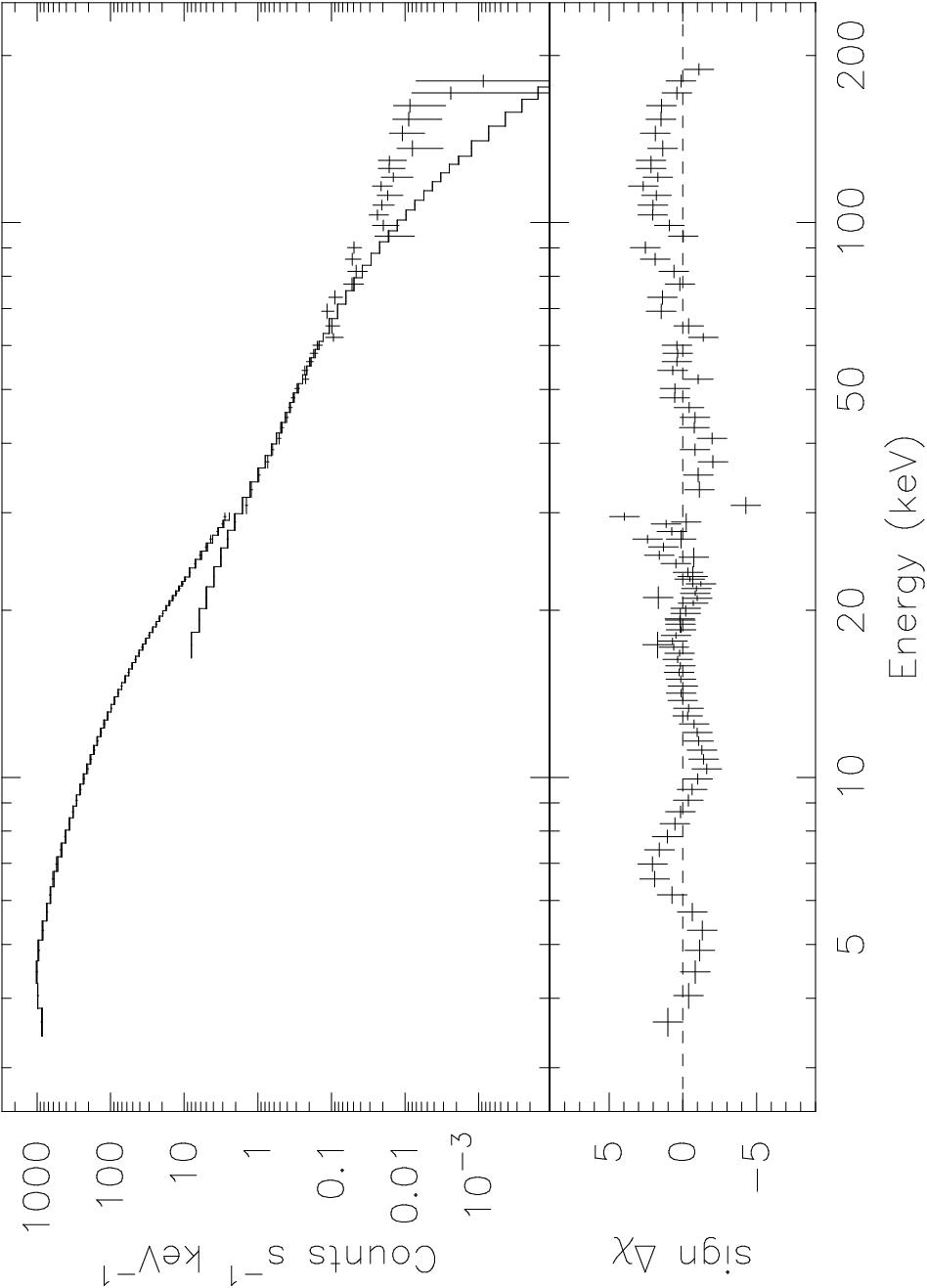}
\caption[]
{ 
The observed count rate spectrum of GRS 1915+105
 obtained from RXTE PCA and HEXTE
on 1999 June 7. 
The error bars are smaller than the symbol sizes in the
low energies. 
A best-fit model
consisting of a disk blackbody and a cutoff power-law is shown as 
histogram and the residuals are shown in the bottom panel 
of the figure. 
}
\end{figure}

The $\chi_3$ state in 1999 June lasted from June 2 to June 7.
The source exhibited a
huge radio flare immediately thereafter (see Naik et al. 2000 for
details of the source properties during this flare episode). We have
selected the RXTE X-ray spectral data obtained on 1999 June 7, just 
prior to this radio flare. The observation PID is 40703-01-17-01.  We have 
made a combined analysis of the data obtained from RXTE PCA (Jahoda et al.
1996) and HEXTE (Rothschild et al. 1998). We have generated the 129 
channel energy spectra from the Standard 2 mode of the PCA.
Standard procedures for
data selection, background estimation and response matrix generation
have been applied. PCA consists of five units (called PCUs) and during
the present observation only three of them (PCU numbers 0, 2, and 3) were
active.  Data from the three units are added together.
For the HEXTE, standard 2 as well as the event mode data have been used.
Systematic errors of 2 \% have been added to the PCA spectral data.
For the present analysis,
we use data from only the HEXTE Cluster 0, which has better spectral response.

\subsection{Spectral analysis}

To understand the systematic errors involved in the data, we have
analyzed the spectra obtained on the Crab nebula. We have also made a systematic
analysis of the Crab data obtained from other missions like Beppo-SAX and 
ASCA. We find that HEXTE Cluster 0 gives satisfactory values of $\chi^2$
when a simultaneous fit to the Crab is made using various missions 
(see Vadawale et al. 2000 for details). Attempt to fit the PCA and HEXTE data
simultaneously gave unacceptable values of $\chi^2$ particularly due to a 
large excess in low energies in PCA (this was noted by Gierlinski et al. 1999).
With the addition of a 2\% systematic error, the 3 -- 30 keV PCA spectrum
of Crab gives statistically acceptable simultaneous fit with the HEXTE
spectrum. We, therefore, have selected only the 3 -- 30 keV spectrum from
the PCA with a systematic error of 2\% for the present analysis.

\begin{figure}[t]
\capt
\vskip 7.0cm
\includegraphics{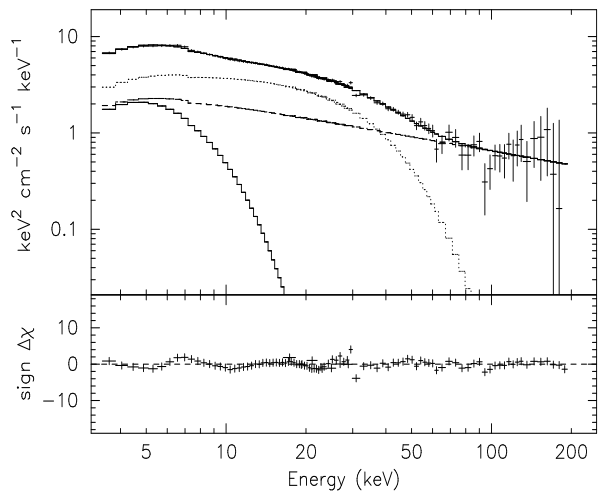}
\caption[]
{ 
The unfolded spectrum (in $\nu F_\nu$ units) shown for the spectrum of
Figure 1. The best-fit model consists of a disk blackbody (continuous
line), a thermal
Compton spectrum (dotted line) and a power-law 
(dashed line).
}
\end{figure}

We have fitted the energy spectrum of the source  using the standard
black hole accretion model (see Muno et al. 1999) consisting
of  disk-blackbody and  power-law with absorption by intervening cold
material parameterized as equivalent Hydrogen column density, N$_H$.
The value of N$_H$ has been kept fixed at 6 $\times$ 10$^{22}$ cm$^{-2}$.
A simultaneous fit to the PCA (in the energy range of 3.5 -- 30 keV)
and HEXTE (in the energy range of 16.5 -- 190 keV)
 data was carried out, keeping
the relative normalization as a free parameter. We find a reasonably 
good fit with a value of reduced $\chi^2$ ($\chi^2_{\nu}$) of
2.2 for 96 degrees of freedom (dof), but the derived value of temperature
was unrealistically high (6.4 keV). Restricting the temperature to
less than 3 keV resulted in a $\chi^2_{\nu}$ of greater than 11.7.
It was also noted by Muno et al. (1999) that the  low hard states
with the radio ``plateau'' emission show very high temperature and low
inner disk radius.
 A cut-off power-law model along with the
disk blackbody gives $\chi^2_{\nu}$ of 1.7. The resultant fitted
spectrum along with the residuals normalized to the errors, are
given in Figure 1. A strong excess above 70 keV is noticeable in the
spectrum.
 We have attempted to model this component as an additional power-law.
It gives an acceptable value of  $\chi^2_{\nu}$ of 1.2. But the
value of the photon index of this additional power-law is 0.5, 
which gives unrealistically
high values of emission above the RXTE range of 200 keV. Hence
we have explored other possible models to fit the spectrum.

 It was shown by Chitnis et al. (1998) that the hard state of Cygnus X-1
can be described by a model consisting of a disk blackbody and  a
thermal Compton spectrum (Sunyaev \& Titarchuk 1980), with
an additional hard component. We have attempted to fit the GRS~1915+105
spectrum using a model consisting of a disk blackbody (diskbb), a thermal-Compton
spectrum (CompST) and a power-law, along with the fixed cold absorption.
A satisfactory fit to the data is obtained with $\chi^2_{\nu}$ value
of 1.2. The resultant unfolded spectrum,
in $\nu F_\nu$ units, along with the residuals, is shown in Figure 2.
The individual model components are also shown in the figure.
The derived parameters are: 1.4$\pm$0.1 keV (disk temperature), 7.0$\pm$1.3
keV (CompST temperature), 6.2$\pm$1.3 (CompST optical depth $\tau$),
and 2.5$\pm$0.4 (power-law photon index). The relative normalization
of HEXTE with respect to PCA is derived to be 0.65$\pm$0.01
(which is the same as the value obtained from the analysis of data
on the Crab nebula). The
quoted errors are nominal 90\% ($\chi^2$ + 4.6 for 2 degrees of freedom).

\subsection{Timing analysis}

For the timing analysis
we have used the  1.5 -- 5.5 keV and 5.5 -- 12.7 keV light
curves (with a time resolution of 16 ms)
from all the 3 active PCUs. 
Power
density spectra (PDS) were generated for 256 bins and then co-added. 
Contribution from Poisson counting statistics has been subtracted.
A  QPO
is clearly detected at a frequency of 2.9 Hz, along with its first harmonic 
and the PDS
is very flat below $\sim$0.5 Hz.  
The centroid frequencies of the QPO and its first harmonics
were determined by fitting Lorentzian functions.
The frequencies derived for the low energy data are 
2.87$\pm$0.01 Hz and 5.36$\pm$0.05 Hz, while that
for the high energy data are 
 2.87$\pm$0.01 Hz and 5.17$\pm$0.08 Hz,  respectively.
The rms variation in the QPO peaks in the low energy data 
are 10\% and 5.3\%, respectively, and the same for the high energy 
data are 16\% and 7.1\%, respectively.

\begin{figure}[t]
\capt
\vskip 6.5cm
\includegraphics{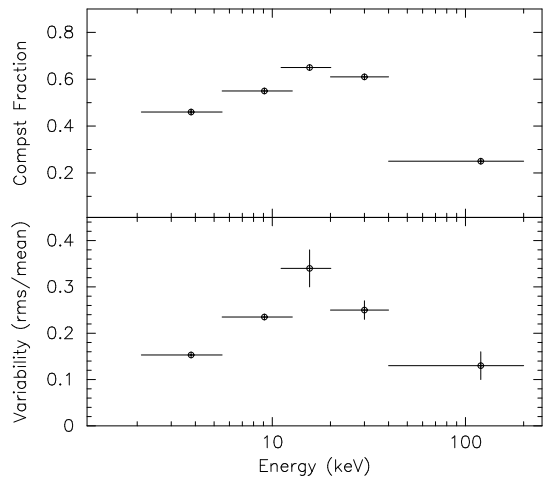}
\caption[]
{ 
The rms variability in 1 -- 4 Hz is plotted as a function of energy
is shown in the bottom panel.
The first two data points are from PCA and the last three are from
HEXTE.  In the top panel the variation in the contribution to the
energy spectrum from the thermal Compton component is shown to
highlight the close similarity between the temporal variability and the
spectral component.
}
\end{figure}

 The increase of the rms in the QPO with energy has been noted earlier 
(Muno et al. 1999), and comparing with Figure 2, it is quite suggestive
that the CompST contribution in the energy spectrum also increases
with energy. To further investigate this, we have taken the event mode data
for HEXTE and generated PDS for 3 energy ranges namely 11.3 -- 20 keV,
20 -- 40 keV and 40 -- 200 keV. The total power in the PDS between
1 Hz -- 4 Hz are calculated. These values are shown as a function of energy
in the bottom panel of Figure 3. We have calculated the relative
contribution of the CompST component to the energy spectrum in 
different energy bins, and these are plotted in the top panel
of Figure 3.
There is a strong correlation between the relative strengths of
the CompST component and the  1 -- 4 Hz rms variability,
suggesting that the CompST 
component is responsible for the variability.

We have also extracted high time resolution light curve of
the three model components.
For this purpose, three energy ranges
are considered: the two energy bands from the PCA as explained earlier, and
11.3 -- 40 keV range from the HEXTE. 
Light curves are  generated at 64 ms time resolution.
The relative contributions of  the three
components are calculated for each of the energy bins
from the fit to the energy spectra.
 By assuming that the time
variability is mainly due to variations in the normalizations of  the
spectral components rather than the spectral parameters, the normalization
constant for each spectral bin are calculated by a $\chi^2$ grid search.
Light curves 
were created for each of the spectral components
with a time resolution of 64 ms. 
PDS were calculated
for the three decomposed spectral components and these
are shown in the right panel of Figure 4.

\begin{figure}[t]
\capt
\vskip 6.5cm
\includegraphics{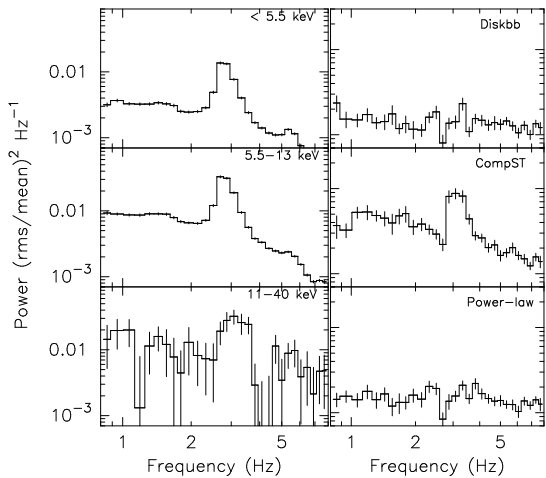}
\caption[]
{ 
The PDS in 3 energy ranges are shown in the left panels of the figure.
In the right panels PDS generated after converting the light curves in 3 
energy channels into light curves in three spectral components are shown.
}
\end{figure}

 The left panel of Figure 4 shows the PDS in the 3 energy bands, where
the QPO is clearly seen. 
The QPO feature is very
prominent for the CompST component  and 
undetectable for the disk blackbody and the
power-law components. This strongly indicates that the CompST component
in the energy spectrum is primarily responsible for the QPO generation.

\section{A  possible model for the QPO}

In the present {\it Letter}, we have shown 
very conclusively that only the Comptonised photons 
participate in QPOs. Chakrabarti \& Manickam (2000) had 
concluded that the emission from the
post-shock
region is responsible for the QPOs, 
by crudely splitting the light
curves into soft photons (0-4) keV and
hard photons (4-13) keV.
Chakrabarti \& Titarchuk (1995)  and Chakrabarti (1997)
while computing  steady state
spectrum from the disk showed that the post-shock region re-emit
photons
intercepted from the Keplerian disk after inverse Comptonisation by the
post-shock flow. They also conjectured
that shock oscillations would produce QPOs in hard radiations and not in
soft radiations.
One way to reconcile the present observation is therefore to imagine
that even when the
oscillation is `on', the two component advective flow (TCAF) model as
proposed by Chakrabarti \& Titarchuk (1995)
and Chakrabarti (1997) remains valid. Oscillation of shocks vary the
puffed-up post-shock region (Molteni,
Sponholz, \&  Chakrabarti, 1996) and intercept variable amount of
soft-photons from the pre-shock
Keplerian region which emits blackbody radiation. Since the fractional
change in the number of
intercepted (and thus emitted) photons in the post-shock region is very
large compared to that
from the Keplerian flow, QPO is strongest in the Comptonised photons and
very nearly absent
in the blackbody photons. Since Keplerian disks do not produce shocks by
definition, it must be surrounded
by the sub-Keplerian flows which does. This was the basis of the
two-component advective flow
models of Chakrabarti \& Titarchuk (1995) and Chakrabarti (1997). What is
new in our present
paper is that we show not {\it all} the hard photons participate in QPOs,
but only those
produced by Comptonisation. This further strengthens the shock
oscillation models.

\section{Discussion and conclusions}

 The aim of the present work is to pin-point the source of QPOs
by making a detailed X-ray spectral analysis. It is  known that
the 0.5 -- 10 Hz QPO is closely related to the hard power-law 
component and  it is used as a tracer to determine the spectral states
(Muno et al. 1999). 
To completely identify the QPO phenomena 
with a particular spectral emission component, a careful fitting
of the X-ray spectrum was required.
 The difficulty of fitting the X-ray spectrum using RXTE PCA has been
highlighted by Gierlinski et al. (1999). We have circumvented this problem
by relying primarily on the HEXTE data (which was validated by the 
Crab analysis) and using PCA at a lower weight by adding a
systematic error of 2\%. Hence the exact parameters of the
low energy components like the black-body temperature etc. need further
confirmation. But, the existence of the various hard X-ray spectral
components is much more secure. The spectral analysis shows that
the model presented in the present work is acceptable.
 We have explored other possible models for the spectra like 
combination of power-law, cut-off power-law etc. along with 
reflection components. We find that disk blackbody with CompST and
power law is preferred by the data compared to all other
models.

The existence of a separate power law is a new feature brought out from the
present analysis. In similar states (radio loud hard state) Muno et al.
(1999) found that the spectrum can be fitted by a disk blackbody and a power law.
We find that a temperature of $>$4 keV and inner disk radius of
$\sim$1 km derived by such a model is quite unrealistic.

 We find that the high energy power-law feature is a generic feature 
in the $\chi_3$ state of the source, when the radio emission is
high (the ``plateau'' state). It is quite possible that this power-law 
X-ray feature is an extension of the  flat-spectrum radio emission
which is identified with a jet of core size 10 AU
(Dhawan et al. 2000).
A systematic search for such a power-law feature and an attempt to
identify this with the radio emission is currently in 
progress (Vadawale et al. 2000).

 The fact that the CompST component in the energy spectrum is responsible for
the QPO generation agrees with the hypothesis that the QPO is generated by
the oscillation of the shock front. This picture explains several of the 
correlations reported in the
literature for the 0.5 -- 10 Hz QPO in GRS~1915+105. 
When the shock front comes closer to the black hole, the QPO frequency should
increase and so should the blackbody flux, as has been observed 
(Markwardt et al.  1999; Chen et al. 1997). When the accretion rate increases, there will be a large number of
soft photons to cool the Compton cloud. This  effectively removes 
the Compton region
in the soft state and   the 0.5 -- 10 Hz QPO is not seen. Recently Reig
et al. (2000) have detected phase lags of both signs between soft and
hard photons during the hard phase of GRS~1915+105. The lags were up to
$\sim$0.3 radians. Such large lags with different signs are difficult
to reconcile with any simple models of Comptonisation. It could be, however,
due to the different shapes of the QPO profile at different energies, which
appear as phase lags.

\begin{acknowledgements}

This research has made use of data obtained through the High Energy
Astrophysics Science Archive Research Center Online Service, provided by the
NASA/Goddard Space Flight Center.

\end{acknowledgements}

{}


\begin{thebibliography}{}


\bibitem[Agrawal et al. 1996]{}
Agrawal, P. C., et al. 1996, IAU Circ. 6488

\bibitem[Belloni et al. 1997]{bell:97}
Belloni, T., et al. 1997, ApJ, 488, L109

\bibitem[Belloni et al. 2000]{bell:00}
Belloni, T., Klein-Wolt, M., Mendez, M., van der Klis, M., \&
 van Paradijs, J. 2000, A\&A,  355, 271

\bibitem[Chakrabarti  (1997)]{chak:97}
Chakrabarti, S. K. 1997, ApJ, 484, 313.

\bibitem[Chakrabarti \& Manickam  (2000)]{chak:00}
Chakrabarti, S. K., \& Manickam, S. G. 2000, ApJ, 531, L41.

\bibitem[Chakrabarti \& Titarchuk (1995)]{chak:95}
Chakrabarti, S. K., \& Titarchuk, L. G. 1995, ApJ, 455, 623. 

\bibitem[Chen et al. 1997]{chen:97}
Chen, X., Swank, J. H., \& Taam, R. E.  1997, ApJ, 477, L41

\bibitem[Chitnis et al. 1998]{chit:98}
Chitnis, V.R., Rao, A.R., \& Agrawal, P.C. 1998, A\&A, 331, 251

\bibitem[]{}
Dhawan, V., Mirabel, I.F., \& Rodriguez, L.F. 2000, ApJ, in press,
astro-ph/0006086

\bibitem[]{}
Fender, R. P., Garrington, S. T., McKay, D. J., Muxlow, T. W. B., 
Pooley, G. G., Spencer, R. E., Stirling, A. M., \& 
Waltman, E. B.  1999, MNRAS, 304, 865

\bibitem[Gerlenski et al. 1999]{gier:99}
Gierlinski, M., Zdziarski, A. A., Poutanen, J, Coppi, P. S.,
 Ebisawa, K., \& Johnson, W. N. 1999, MNRAS, 309, 496 


\bibitem[Jahoda et al. 1996]{jaho:96}
Jahoda, K., et al. 1996, SPIE, 2808, 59

\bibitem[Markwardt et al. 1999]{mark:99}
Markwardt, C.B., Swank, J.H., \& Taam, R.E. 1999, ApJ, 513, L37

\bibitem[Mirabel \& Rodriguez 1999]{mira:99} 
Mirabel, I.F. \& Rodriguez, L.F. 1999,  ARA\&A, 37, 409

\bibitem[Morgan \& Remillard  1996]{morg:96}
Morgan, E. H. \& Remillard, R. A. 1996, IAU Circ., 6392

\bibitem[Molteni et al. 1996]{molt:96}
Molteni, D., Sponholz, H., \& Chakrabarti, S. K. 1996, ApJ, 457, 805.

\bibitem[Morgan et al. 1997]{morg:97}
Morgan, E. H., Remillard, R. A., \& Greiner, J. 1997, ApJ, 482, 993

\bibitem[Muno et al. 1999]{muno:97}
Muno, M.P., Morgan, E.H., \&  Remillard, R. A. 1999, ApJ, 527, 321.

\bibitem[]{}
Naik, S. et al. 2000, ApJ, submitted

\bibitem[Paul et al. 1998]{paul:98}
Paul, B., et al. 1998, A\&AS, 128, 145

\bibitem[]{}
Reig, P., Belloni, T., van der Klis, M., Mendez, M., Kylafis, N., 
\& Ford, E.C. 2000, astro-ph/0001134

\bibitem[]{}
Rothschild, R.E., et al. 1998, ApJ, 496, 538

\bibitem[]{}
Sunyaev, E.A., \& Titarchuk, L. 1980, A\&A, 86, 121

\bibitem[Trudolyubov  et al. 1999a]{trud:99a}
Trudolyubov, S., Churazov, E., \& Gilfanov, M. 1999a, A\&A, 351, L15

\bibitem[Trudolyubov  et al. 1999b]{trud:99b}
Trudolyubov, S., Churazov, E., \& Gilfanov, M. 1999b, Ast. L., 25, 718

\bibitem[]{}
Vadawale, S.V., et al., 2000, in preparation

\bibitem[Yadav et al. 1999]{yadav:99}
Yadav, J. S., Rao, A. R., Agrawal, P. C., Paul, B., Seetha, S., \&
Kasturirangan, K. 1999, ApJ, 517, 935

\end{thebibliography}
\end{document}